\documentclass[
 superscriptaddress,
 nofootinbib,
 amsmath,
 amsfonts,
 amssymb,
 aps,
 prd,
 floatfix,
 notitlepage,
 twocolumn
]{revtex4-2}

\usepackage{aas_macros} 
\usepackage{graphicx}
\usepackage{dcolumn}
\usepackage{float}
\usepackage[top=1.0in,bottom=1.0in,left=0.8in,right=0.8in]{geometry}
\usepackage[mathscr]{euscript}
\usepackage{verbatim}
\usepackage[colorlinks=true, allcolors=blue]{hyperref}
\usepackage[capitalise]{cleveref}
\usepackage{soul}

\def\AxioNyx{\textsc{AxioNyx} }
\def\AxioNyxns{\textsc{AxioNyx}}

\newcommand{\ii}{{\rm i}}

\newcommand{\rb}{\mathbf{r}}

\begin{document}

\title{Multifield Ultralight Dark Matter}

\author{Mateja~Gosenca}
\email{mateja.gosenca@univie.ac.at}
\affiliation{Faculty of Physics, University of Vienna, Boltzmanngasse 5, 1090 Vienna, Austria}

\author{Andrew~Eberhardt}
\affiliation{Department of Physics, Stanford University, Stanford, CA 94305, USA}
\affiliation{Kavli Institute for Particle Astrophysics and Cosmology, Menlo Park, California 94025, USA}
\affiliation{SLAC National Accelerator Laboratory, 2575 Sand Hill Rd, Menlo Park, California 94025, USA}

\author{Yourong~Wang}
\affiliation{Department of Physics,
University of Auckland,
Private Bag 92019,
Auckland, New Zealand}

\author{Benedikt~Eggemeier}
\affiliation{
 Institut f\"ur Astrophysik, Georg-August-Universit\"at G\"ottingen, D-37077 G\"ottingen, Germany
}

\author{Emily~Kendall}
\affiliation{Department of Physics,
University of Auckland,
Private Bag 92019,
Auckland, New Zealand}

\author{J. Luna~Zagorac}
\affiliation{Department of Physics, Yale University, New Haven, CT 06520, USA}
\affiliation{
Perimeter Institute for Theoretical Physics,
31 Caroline St. N., 
Waterloo, ON N2L2Y5, Canada}

\author{Richard~Easther}
\email{r.easther@auckland.ac.nz}
\affiliation{Department of Physics,
University of Auckland,
Private Bag 92019,
Auckland, New Zealand}

\date{\today} 
\begin{abstract}
\noindent 
\noindent Ultralight dark matter (ULDM) is usually taken to be a single scalar field. Here we explore the possibility that ULDM consists of $N$ light scalar fields with only gravitational interactions. This configuration is more consistent with the underlying particle physics motivations for these scenarios than a single ultralight field. ULDM halos have a characteristic granular structure that increases stellar velocity dispersion and can be used as observational constraints on ULDM models.
In multifield simulations, we find that inside a halo the amplitude of the total density fluctuations decreases as $1/\sqrt{N}$ and that the fields do not become significantly correlated over cosmological timescales.
Smoother halos heat stellar orbits less efficiently, reducing the velocity dispersion relative to the single field case and thus weakening the observational constraints on the field mass.  Analytically, we show that for $N$ equal-mass fields with mass $m$ the ULDM contribution to the stellar velocity dispersion scales as $1/(N m^3)$. 
Lighter fields heat the most efficiently and if the smallest mass $m_L$ is significantly  below the other field masses the dispersion scales as $1/(N^2 m_L^3)$.
\end{abstract}
\maketitle

\section{Introduction}

Axion Dark Matter is a promising dark matter candidate. Its ultralight limit (ULDM), also called fuzzy dark matter (FDM), exhibits unique interference phenomena on galactic scales, while still behaving like cold dark matter (CDM) on scales much larger than the de~Broglie wavelength.

CDM has had spectacular success accounting for the 
formation of structures in the Universe. In particular, it predicts  the  features of the cosmic web
\cite{1983MNRAS.204..891K,1985ApJ...292..371D} revealed by galaxy surveys and the 
properties of the anisotropies in the microwave background \cite{1982ApJ...263L...1P}.  However, at small scales many details of galactic dynamics remain unclear.  Specific issues include the missing satellites \cite{klypin1999, Moore1999}, core-cusp \cite{Navarro:1995iw}, and too-big-to-fail \cite{Boylan-Kolchin2011} problems associated with small-scale structure formation in CDM, as reviewed in Ref. \cite{Bullock2017}. It is possible that these problems will be solved by improving survey sensitivity \cite{Nadler_2021, Drlica-Wagner_2020,Kim_2018},  baryonic physics~\cite{d_onghia_2010} 
or even the breakdown of Newtonian dynamics~\cite{2021PhRvL.127p1302S} at galactic scales.  Another option is that the dark matter has more complicated small-scale dynamics than  predicted by CDM alone and  ULDM  is a widely-studied scenario in this category.  Various extensions to this simple model have been  considered, including mixed dark matter scenarios~\cite{Schwabe_2020} and self-interacting scalar fields~\cite{Barranco:2010ib, Shapiro:2021hjp,Glennon:2022huu}.

Given that the characteristic signatures of ULDM require wavelike effects to be visible at subgalactic scales but must not completely preclude the existence of small-scale structure, a particle mass of 
$10^{-22} - 10^{-18} \, \mathrm{eV}$
is favored~\cite{Hu:2000ke,Chen:2016unw}. However, recent studies have  
put limits on much of this parameter space. 
A non-exhaustive list of the phenomena that can provide constraints includes the Lyman-$\alpha$ forest~\cite{Armengaud:2017nkf, Irsic:2017yje, Nori:2018pka, Rogers_2021}, the galactic subhalo mass function~\cite{Nadler_2021, Schutz2020}, stellar dispersion of ultra-faint dwarfs~\cite{Marsh:2018zyw,Dalal2022}, galactic density profiles~\cite{Bar2018,Bar_2022, Zoutendijk:2021kee}, and Milky Way satellites~\cite{Safarzadeh_2020}. A recent review is given in Ref.~\cite{Hui_2021}. 

Critically, almost all treatments of ULDM assume the presence of a single ultralight field. However, much of the motivation for ULDM comes from string-theoretic approaches to high-energy physics. These typically support many axion-like fields rather than just one, as the axions are associated with (non-equivalent) closed 2-cycles (two-dimensional submanifolds of a larger manifold that cannot be smoothly contracted to a point)  of the  Calabi-Yau manifold that sets the topology of the compact dimensions. 
Calabi-Yau manifolds can contain many closed cycles -- numbers in the hundreds are seen as typical~\cite{Hui:2016ltb}.
From this perspective, multiple axion-like fields are actually a more reasonable assumption than 
a single ultralight field.
It is expected that their masses are distributed broadly uniformly on a logarithmic scale~\cite{Arvanitaki:2009fg}, so several axion-like species may be expected per decade of mass between $10^{-33}\,\mathrm{eV}$ and $10^8\,\mathrm{eV}$. However, the precise nature of the distribution of masses is not tightly constrained, and it is possible that it may be characterized by densely populated mass ranges separated by larger gaps. 

Single-field UDLM has rich dynamics on subgalactic scales and multifield ULDM is potentially substantially more complex. 
In this paper, we explore this scenario with a specific focus on the heating of stellar orbits in ULDM halos. ULDM halos have granular structures on scales similar to the de~Broglie length of the underlying field. 
Orbiting stars are perturbed as they interact with the gravitational potential of these structures, heating them relative to their motion in a smooth background. This effect is enhanced when the ULDM mass is small -- in this limit the granules are larger and more massive and the resulting kicks to stellar velocities are more substantial. Consequently, the observed velocity dispersions of stars in dwarf halos leads to lower bounds on the ULDM mass. However, when several uncorrelated fields are present, their granules overlap leading to a smoother density field which reduces the heating.

Previous studies of multifield scenarios include treatments of large-scale structure~\cite{Tellez-Tovar:2021mge} and rotation-curve~\cite{Street:2022nib} constraints on two-field scenarios.  Guo {\em et al.}~\cite{2021JCAP...10..028G} consider the dynamics of two-field condensates with possible nonlinear mutual and self-interactions, while Luu {\em et al.}~\cite{2020PDU....3000636L} consider the properties of nested solitons in multifield scenarios.  Very recently, Huang {\em et al.}~\cite{Huang:2022ffc} have performed several two-field cosmological simulations which showed that the 
central soliton can be substantially modified in this scenario, potentially in ways that better match observations. Finally, Amin {\em et~al.}~\cite{Amin:2022pzv} looked at the related case of ultralight vector dark matter which can mimic three equal-mass scalar fields in the non-interacting case and  note the consequent reduction in stellar heating.  
 
In this work, we use the superposition of eigenstates to construct self-consistent ULDM halos. We then numerically evolve these halos in order to analyze how the granularity in the outer regions depends upon the number of distinct ULDM fields present. We consider up to four ULDM fields, in both equal-mass and mixed-mass scenarios. Visually, halos appear smoother with increasing numbers of fields and we explicitly confirm this by calculating the two-point correlation function.
The granules in each field are effectively a system of oscillators, coupled through the gravitational potential so there is the possibility
for their motion to become correlated over time. If this 
happened,
it would undermine the suppression of the heating in multifield models and we find that while a small correlation arises it does not grow with time over 5~Gyr for typical  axion masses.

We combine the results of these simulations with the analytic approximations
of multifield stellar heating.
In particular, 
the predicted stellar dispersion decreases as $1/(N m^3)$ for $N$ fields
in the equal-mass case of field with mass $m$.
In the multi-mass case, the lightest field dominates the heating since the 
strength of the heating decreases with the third power of
the particle mass. If the lightest field with mass $m_L$ is 
sufficiently light compared to the 
second lightest field the resulting dispersion scales as $1/(N^2 m_L^3)$. 
 
The paper is organized as follows. 
In \cref{sec:Method} we outline our conventions and methodology, describe how initial conditions are set up, and define statistical parameters used to analyze the results. 
In \cref{sec:simResults} we discuss the results of multifield simulations with both equal or different particle masses. 
In \cref{sec:stellarDispConstr} we present an analytic estimate of how the velocity dispersion induced by multifield ULDM scales with the number of fields. We discuss our results in \cref{sec:Conclusions}.

\section{Methodology}
\label{sec:Method} 

We assume that dark matter is composed of $N$ independent classical, real scalar fields which only interact gravitationally with each other. In general, the evolution of these fields is governed by the Klein-Gordon-Einstein equations. Considering the nonrelativistic limit of the Klein-Gordon equation relevant for structure formation, the evolution equation in an expanding Friedmann-Lema\^itre-Robertson-Walker universe reduces to the Schr{\" o}dinger equation~\cite{Ruffini1969,Nambu1990}
\begin{equation}
  i\hbar \frac{ \partial}{\partial t}\Psi_i = -\frac{\hbar^2}{2m_ia^2}\nabla^2\Psi_i + m_i\Psi_i\Phi \, 
  \end{equation}
where $m_i$ is the mass of the $i$-th field and $a$ is the cosmological scale factor. In this limit, the $N$ scalar fields are represented by their corresponding complex wavefunctions $\Psi_i$. This wavelike matter responds to 
the total gravitational potential $\Phi$, 
which is the solution to the 
Poisson equation. In comoving units, 
\begin{equation}
    \nabla^2\Phi = \frac{4\pi G}{a}(\rho-\bar\rho)\,,\quad \mathrm{with}\quad\rho = \sum_{i=1}^N m_i |\Psi_i|^2\,,
\end{equation}
where the source term for the field is the difference between the total local density $\rho$ and the average density $\bar\rho$.  As usual, Planck's constant is represented by $\hbar$ and Newton's constant by $G$.   

We can set $a=1$  since we focus on individual halos that are decoupled from the Hubble flow. 
In the ultralight limit, the particle mass $m$ is often expressed as $m = m_{22} \times 10^{-22}\,{\rm eV}$, and we  adopt this notation.  The
fraction of the total mass 
contained in
the $i$-th field is denoted by $c_i$, with $\sum_i^N c_i =1$. Since the fields do not 
exchange mass
$c_i$ are constants and we work with $c_i = 1/N$ in our simulations. We consider cases with $N$ fields with identical masses and multi-mass scenarios where the masses are distinct. 

Single-field ULDM halos have an excited 
central soliton  ~\cite{Veltmaat_2018,Eggemeier:2019jsu,Veltmaat:2019hou,Schwabe_2020} that oscillates with a frequency given by its quasi-normal modes~\cite{Guzman_2004}.
This soliton is surrounded by a halo composed of interfering excited states which yield
approximately spherical density fluctuations, usually called halo granules. The size of these granules is roughly given by the de~Broglie wavelength
\begin{equation}
    \label{debroglie}
 \lambda_{{\rm dB}} = \frac{\hbar}{m v} \,,
\end{equation}
where $v$ denotes the local velocity of the field inside the halo.
This length is of course different for each field 
if the fields have different particle masses.

We perform numerical simulations with a version of \AxioNyx~\cite{Schwabe_2020} modified to evolve $N$ 
fields. \AxioNyx features adaptive mesh refinement~(AMR) which allows computing power to be focused on regions of interest -- in this case the halo center -- while the outer regions are evolved at lower resolution. In the multi-mass case, the density-based refinement condition is taken from the heaviest field in the simulation which has the smallest de~Broglie wavelength. All fields are evolved on the same grid structure. In the equal-mass cases, the refinement criterion is evaluated with respect to a single, arbitrarily chosen field. 

The Schr\"odinger equation is solved with a spectral method on the root grid with periodic boundaries. In refined regions we use finite differencing with aperiodic boundary conditions and an appropriately subcycled timestep. The Poisson equation is solved using the multigrid Gauss-Seidel redblack solver implemented within \textsc{Nyx}~\cite{Almgren2013}. All simulations are initialized on a $(100\,\mathrm{kpc})^3$ box and have a $128^3$ root grid resolution and three levels of refinement in the center. Spatial resolution is increased by a factor of two on each refinement level.

\subsection{Halo Construction} \label{sec:Halo}

We generate multifield halos directly by adapting 
the eigenmode method described in Ref.~\cite{Yavetz:2021pbc}  in which the initial configuration is constructed by decomposing the halo density profile into radial eigenfunctions (see also Ref.~\cite{Zagorac:2021qxq}) and multiplying each of them with a random phase. 
We extend this to the multifield case by using the same radial eigenfunctions for the density (appropriately scaled by $c_i$), but a different set of random phases for each field.
See \cref{app:ic} for more details. 

We assume that each field has a central soliton and a surrounding NFW halo 
with a combined density profile
\begin{equation}
\label{eq:combined_profile}
    \rho(r) = 
    \begin{cases}
      \rho_{\rm sol} & \text{if $\rho_{\rm sol} > \rho_{\rm NFW}$}\\
      \rho_{\rm NFW} \: e^{[-(r/r_{\rm vir})^2/2]}& \text{otherwise}.\\
    \end{cases}      
\end{equation}
The exponential term suppresses the density outside the virial radius $r_{\rm vir}$ to minimize  interactions at the  periodic boundaries of the box.

The NFW profile is given by~\cite{Navarro:1995iw}
\begin{equation}
    \rho_{\rm NFW} (r) = \frac{\rho_0}{r/r_s\left(1+r/r_s \right)^2}\,
\end{equation} 
where $r_s$ is the scale radius. According to convention, we define the virial radius as the radius at which the average density in the enclosed sphere is 200 times the critical density of the Universe, which also sets the virial mass inside this sphere, $M_{\rm vir}$.
Finally, $\rho_0$ is determined by integrating this profile up to the virial radius.

For a single field, the central solitonic core is well-described by a fitting formula~\cite{Schive2014_PRL}
\begin{align}
    \rho_\mathrm{sol}(r) = \frac{\rho_s}{\left(1 + 0.091\left(r/r_c\right)^2\right)^{8}}\,,
    \label{eq:sol_fit}
\end{align}
where $r_c$ denotes the core radius where the density is half of its central value and
$\rho_s = 1.9 \times 10^{9} \, m_{22}{}^{-2} \, ( r_c / \mathrm{kpc})^{-4}\,{\rm M}_{\odot}\,{\rm kpc}^{-3}$. 

We use the virial halo mass $M_{\rm vir}$ to determine the core radius $r_{\rm c}$ of the soliton applying the core-halo relation introduced in~\cite{Schive2014_PRL}, 
\begin{equation}
r_\mathrm{c} = 1.6 \times \frac{1}{m_{22}}  \left(\frac{M_\mathrm{vir} }{10^9 M_\odot} \right)^{-1/3} \mathrm{kpc}.
\end{equation}
This relation has become contested
with several works producing different scaling relationships showing the sensitivity of results on numerical methods rather than physics, or even arguing against a universal core-halo mass relation (see Refs.~\cite{Kendall2020,Chan2022,Taruya2022,Zagorac2022} and works discussed therein). Nevertheless, we assume that it is a sensible estimate of core size in the multifield case. 
Finally, we are free to choose the NFW scale radius  in the range between $r_c$ and the virial radius and this choice does not significantly affect the dynamics of the halo. 

For the sake of definiteness, we choose the dwarf galaxy Eridanus II as a template for the halo as it has been widely used to test ULDM. Eridanus II has a half-light radius $r_\mathrm{1/2} = 300\,\mathrm{pc}$~\cite{2016crnojevic} and half-light mass $M_\mathrm{1/2} = 1.2 \times 10^{7}\,M_\odot$~\cite{DES:2016vji}.
According to Ref.~\cite{Kravtsov:2012jn}, the half-light radius and the virial radius are related by $r_\mathrm{1/2} = 0.015 \: r_\mathrm{vir}$, corresponding to $r_\mathrm{vir} = 20\,\mathrm{kpc}$ and $M_\mathrm{vir} =4\pi/3\times 200 \: \bar{\rho} \: r_\mathrm{vir}^3 \simeq 3\times 10^8\,M_\odot$. This results in the core radius $r_c \simeq 0.5$~kpc and we choose $r_s = 2\,\mathrm{kpc}$ for the scale radius.

\begin{figure*}[t]
    \includegraphics[width=\textwidth]{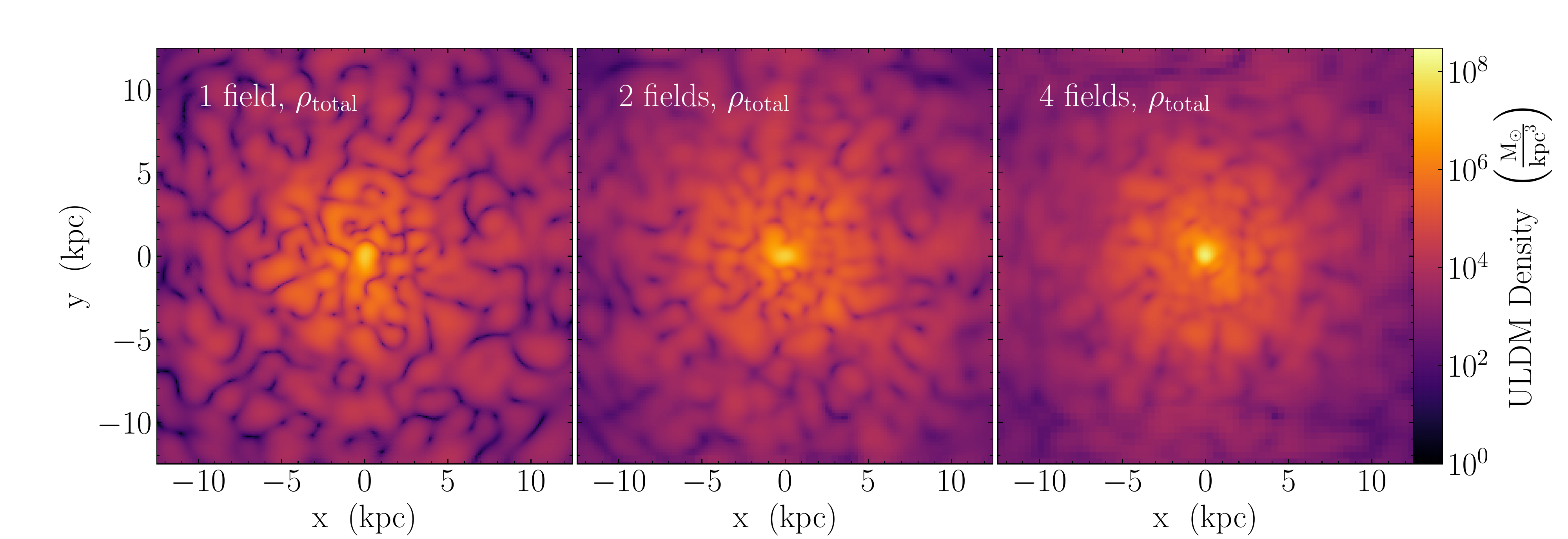}
    \caption{Total density around the center of the halo for the simulations involving one field, two fields, and four fields. The total density becomes progressively smoother as the number of fields increases.}
    \label{fig:total_dens}
\end{figure*}

\subsection{Fluctuation Statistics}
We define the two-point correlation function 
\begin{equation}
\begin{split}
    \xi(d) 
    & = \left\langle \delta({\bf x}) \delta({\bf x+d})\right\rangle \\
    & =\frac{1}{V} \int  \delta({\bf x}) \delta({\bf x+d})\,{\rm d}^3 x
\end{split}
\end{equation}
to quantify the smoothness of the density field.  
We first obtain the spherically-averaged density profiles $\bar{\rho}(x)$ around the highest-density point which only depends on the distance $x$ from that point. The overdensity field is 
\begin{equation}
    \delta({\bf x})= \frac{\rho({\bf x}) - \bar{\rho}(x)}{\bar{\rho}(x)}\,.
\end{equation}
This quantity is sampled at $n$ random points with coordinates ${\bf x}$ inside a spherical domain with radius $x_{\rm max}$.

We multiply the overdensity in all $n(n-1)/2$ pairs of points where two points are separated by a vector ${\bf d}$ and bin them according to the distance $d = |${\bf d}$|$ between the pair,
\begin{equation}
    \xi(d_k) 
    =\frac{1}{n_k}\sum_{i=0}^{n}\sum_{j=i+1}^{n} \delta({\bf x}_i)\delta({\bf x}_j)W_k(|{\bf x}_i -{\bf x}_j|)\,,
\end{equation}
where $n_k$ is the number of pairs of points in each bin, and we take the mean value to give the result for the bin. The window function $W_k(|{\bf x}_i -{\bf x}_j|) =1$ if the distance between ${\bf x}_i$ and ${\bf x}_j$ falls into the $d_k$ bin, and zero otherwise.
We verify that the sample size $n$ is large enough to ensure convergence. 

The fields are initialized to be completely uncorrelated, but correlation can in principle grow as the system evolves. In particular, an overdensity in one field induces a local gravitational well which  
influences the
dynamics of the other fields. This could in principle create a correlation between the fields which would grow over time. To assess this possibility we introduce the reduced one-point covariance or correlation parameter 
\begin{equation}
\label{eq:1p-cov}
  \zeta({\bf x}) = 
  \frac{\langle \delta_1({\bf x}) \delta_2({\bf x}) \rangle}{\sqrt{\langle \delta_1({\bf x})^2 \rangle} \sqrt{\langle \delta_2({\bf x})^2 \rangle}}\,,
\end{equation}
where  $\langle \cdots \rangle$ again denotes a spatial average. 
For completely correlated fields $\zeta=1$, for anti-correlated fields  $\zeta=-1$, and $\zeta=0$ for entirely uncorrelated fields.  

\section{Simulations} 
\label{sec:simResults}

\subsection{Equal-mass}\label{sec:simulations_equal_mass}

We performed simulations with ULDM mass $m_{22} = 5$ and one, two, and four ULDM fields. Each field in the multifield simulations is initialized with the same radial eigenfunctions, but different random phases. The granules of the different fields are thus initially uncorrelated.  
We evolve this system through $\mathcal{O}(30)$ oscillation periods, or roughly $5\,\mathrm{Gyr}$. 
Transients associated with the relaxation of the initial state decay over the first two or three oscillation times. 

The density on a slice through the center of the simulations is shown in \cref{fig:total_dens}, at a representative time ($1.6\,\mathrm{Gyr}$). As the number of fields increases, the  solitonic core retains its shape
because there is a central overdensity in each of the constituent fields. At the same time, the granular overdensities in the surrounding halo are visibly smoothed out.

\begin{figure*}[tb]
  \includegraphics[width=1.0\textwidth]{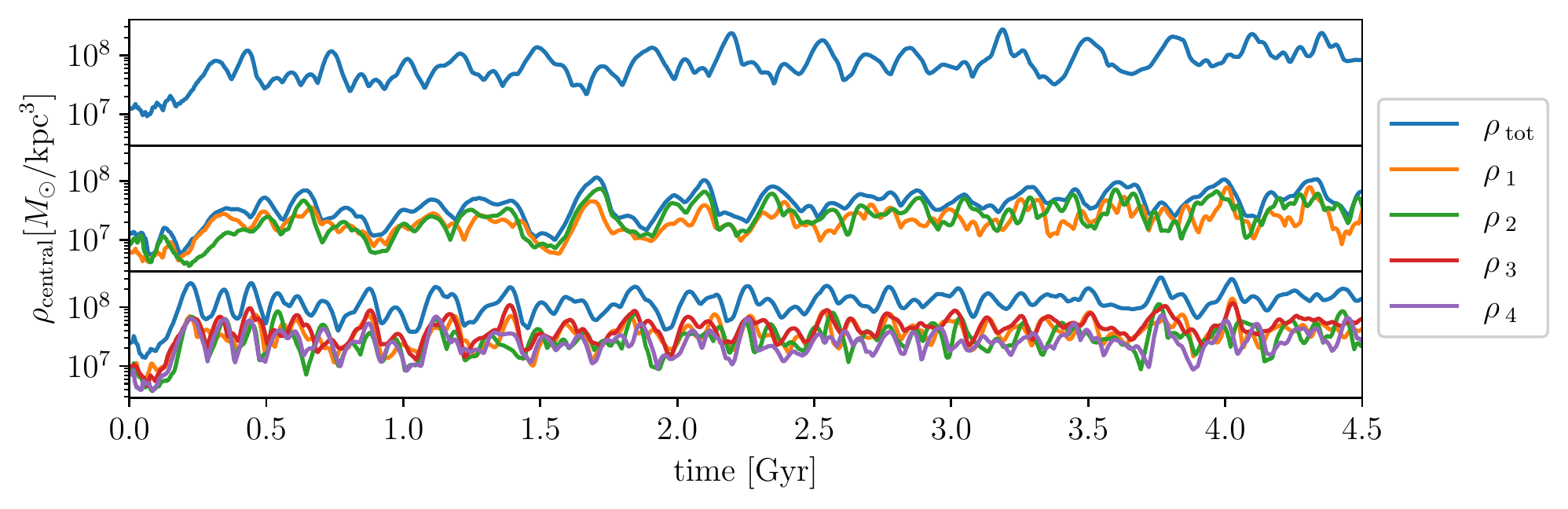}
  \caption{The density in the center of the soliton for the simulations with only one field (top), two fields (middle), and four fields (bottom). We show densities of individual constituents $\rho_i$, as well as the total density $\rho_\mathrm{tot}$.}
  \label{fig:eqmass_dens}
\end{figure*}

\cref{fig:eqmass_dens} shows the evolution of the maximum density, which corresponds to the central density of the soliton, as a function of time. 
In all cases we see oscillations in the  solitonic core~\cite{Veltmaat_2018, Guzman_2004}, along with an initial transient. In the multifield scenarios, the central oscillations in constituent fields are synchronized.
While they initially overlap owing to our choice of initial conditions, there is also a clear synchronization over time, presumably maintained by their mutual gravitational coupling. 

\begin{figure}[tb]
    \includegraphics[width=0.5\textwidth]{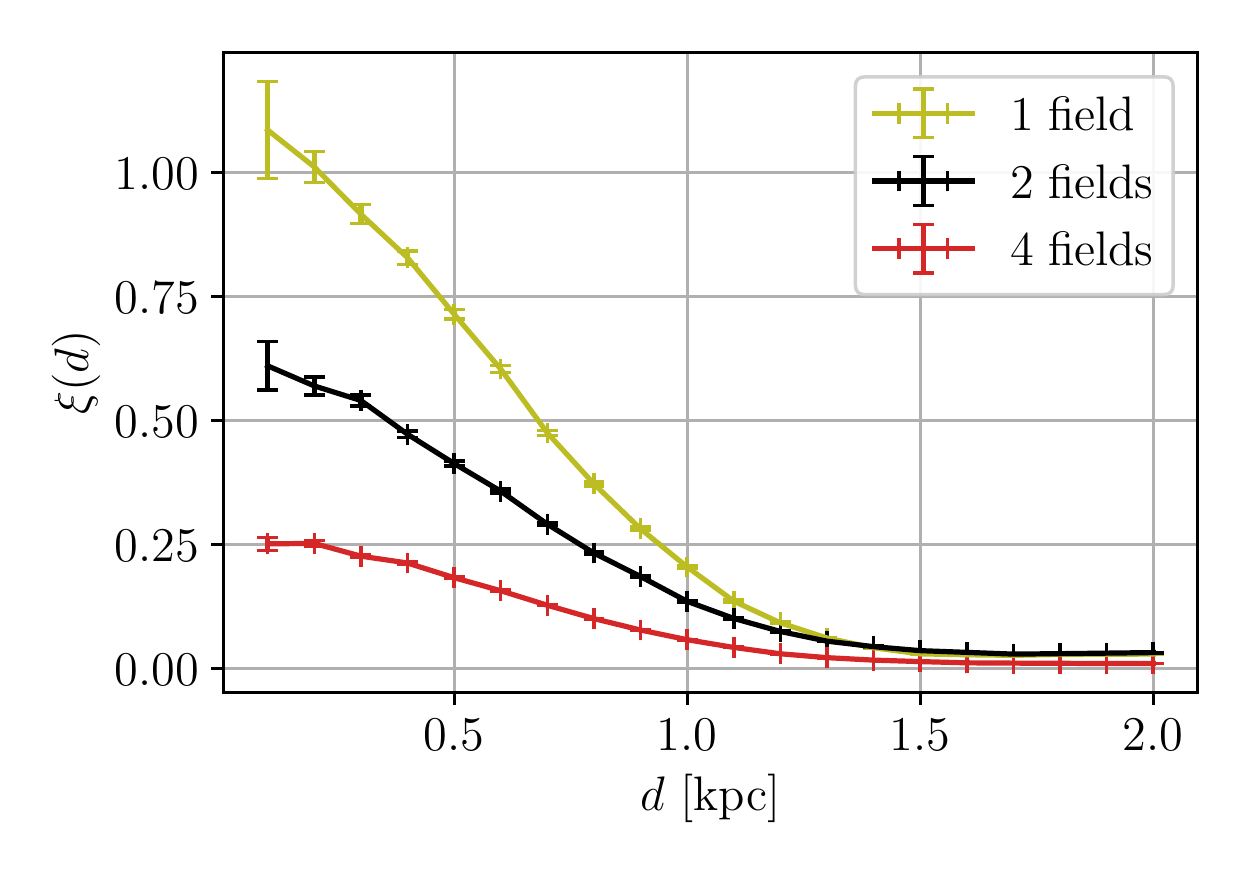}
  \caption{Two-point correlation function of the total overdensity for the three equal-mass simulations at an arbitrary time. We see that $\xi \sim 1/N$ and that the fields are fully uncorrelated at scales much larger than the de~Broglie length.}
  \label{fig:124nfields_2pcf} 
\end{figure}

The two-point correlation function for the equal-mass simulations 
is shown in \cref{fig:124nfields_2pcf}. 
It is calculated at an arbitrary time during the simulations.
The value of $\xi(r)$ is close to zero for scales much larger than the de~Broglie wavelength. Critically, $\xi(r)$ scales as $1/N$, from which we can infer that the amplitude of the overdensity decreases in proportion to $\sqrt{N}$, i.e. in the multifield case $\delta(x) \rightarrow \delta(x) / \sqrt{N}$.
To compute $\xi$ we sampled the overdensity in 
random points from a spherical domain centered on the center of the simulation box with a radius of $10$~kpc. While the solitons 
undergo a random walk~\cite{Schive:2019rrw,Li2021} around their 
initial position, they are rarely found  more than $1.0$~kpc from the center of the box, and  usually within $0.5$~kpc. This permits us to center our spherical domain on the box rather than the location of the soliton.

\begin{figure}[tb]
    \includegraphics[width=0.48\textwidth]{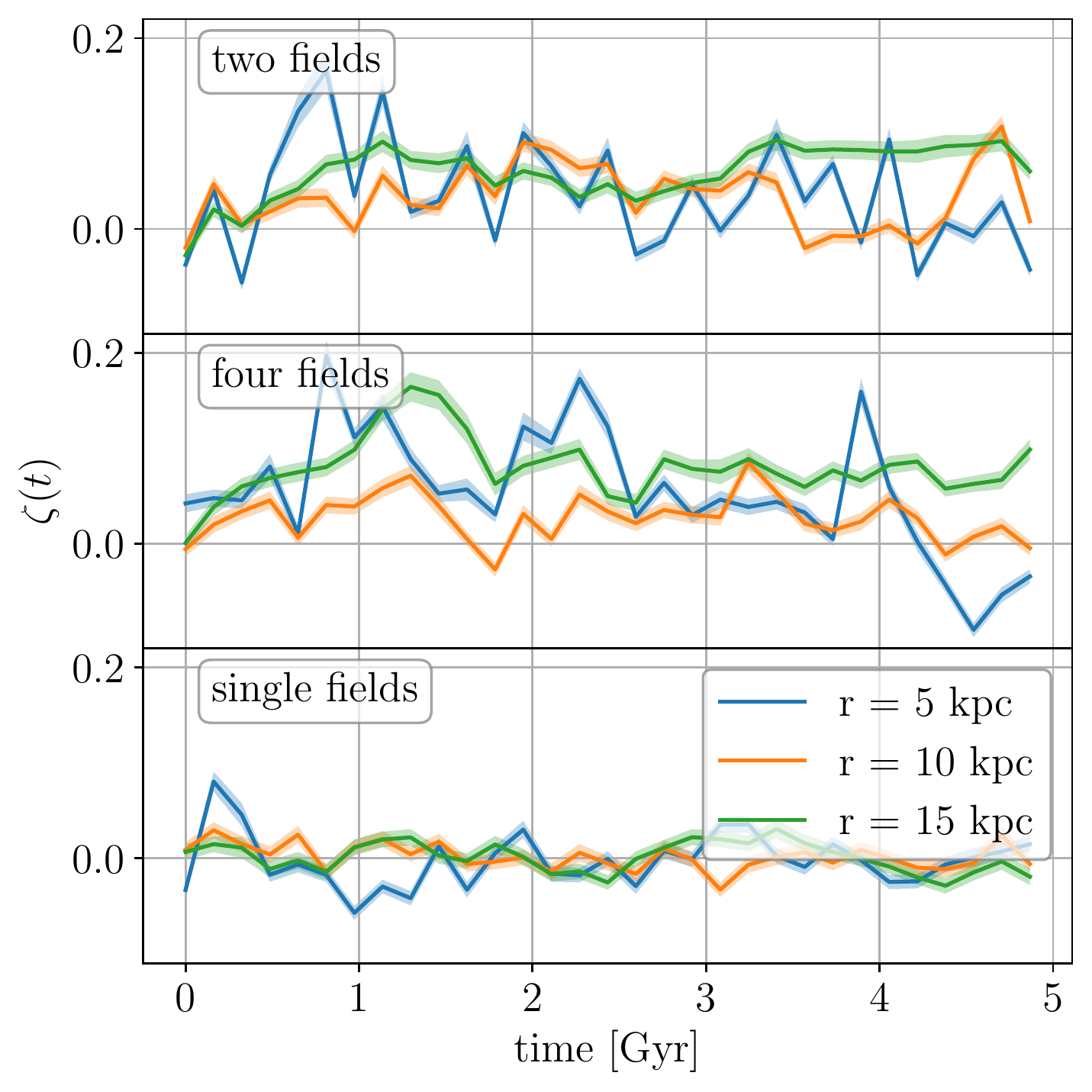}
    \caption{The value of the correlation parameter for a combination of two constituent fields in simulations with two (top panel) and four (middle panel) fields. In the four-field case, we show only the correlation between two arbitrarily selected fields; other combinations give very similar results. For reference, we show the same statistical measure for two unrelated one-field simulations (bottom panel). The uncertainty of $\zeta$ at each time is captured by the standard error which is shown as the shaded area around each line.}
    \label{fig:cov_all}
\end{figure}

If the granules developed spatial correlation over time, the suppression of the overdensity observed in \cref{fig:124nfields_2pcf} would  diminish. 
We use the correlation parameter $\zeta$ (see \cref{eq:1p-cov}) to test whether such correlation develops, and show $\zeta$  
for the two-field and four-field simulations in the top two panels of  \cref{fig:cov_all}. At the start of each simulation, constituent fields are initialized with uncorrelated fluctuations which manifests as $\zeta (t=0)$ being around zero. We present results for cut-off radii of 5, 10, and 15~kpc. 
A small centrally-positioned sphere of radius $3\,r_c$ was excised from the sampling in order to exclude the solitonic core where a positive correlation is to be expected. In practice, however, since the volume of this sphere is tiny compared to the rest of the sampling domain, including this region would not significantly change our results. 

In each of the multifield runs $\zeta$ is initially close to zero, but a small, positive correlation between the fields develops in the first 0.5 billion years of simulated evolution. However, this growth does not continue and the correlation remains small on timescales akin to the present age of the Universe. Moreover, while we have not made an exhaustive study we do see that the correlation is roughly similar in both two- and four-field scenarios. 

For comparison, we  show the correlation between two separate single-field simulations in the bottom panel of \cref{fig:cov_all}. In this case, as expected, $\zeta$ fluctuates around zero, demonstrating that the small positive value of $\zeta(t)$ shown for the two- and four-field runs is indeed physical, rather than a computational artefact. 

\begin{figure}[tb]
  \includegraphics[width=0.49\textwidth]{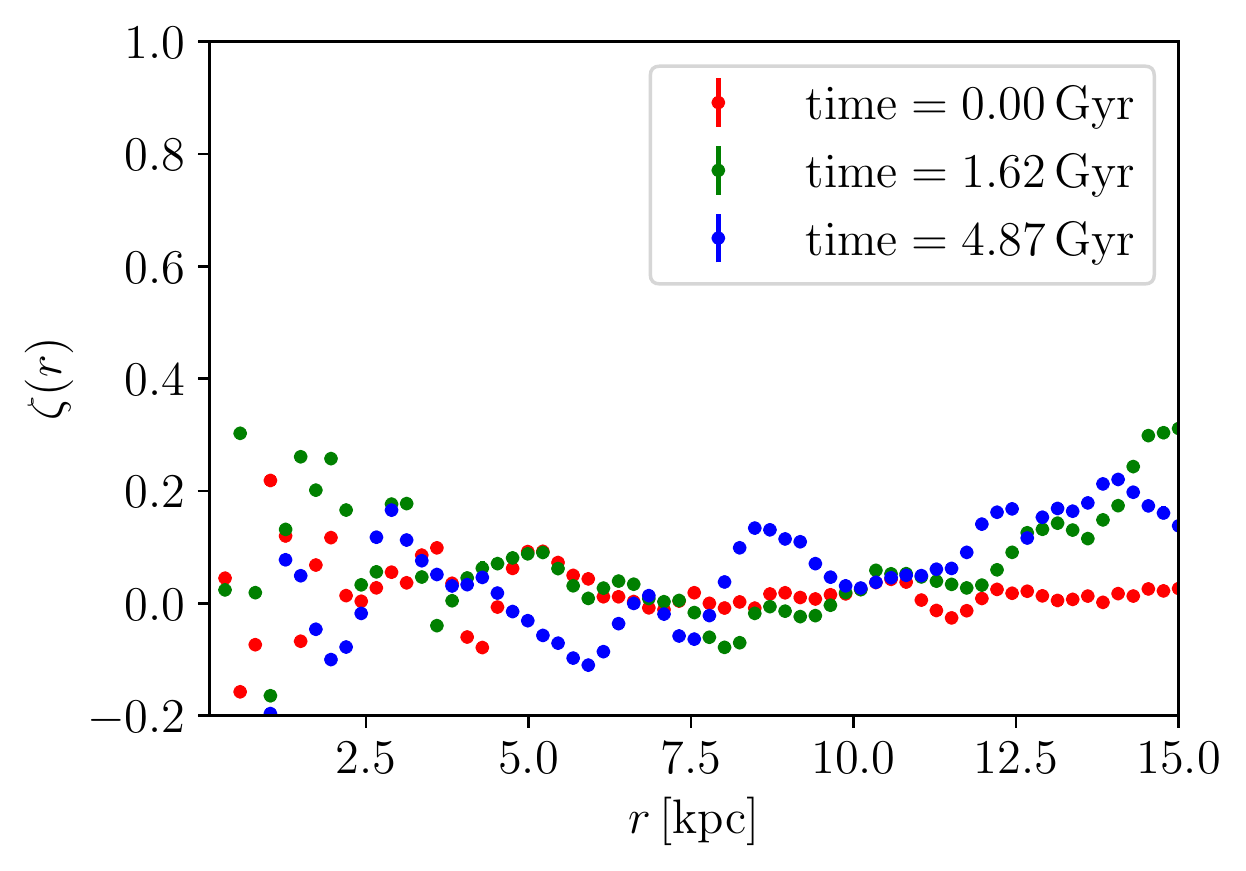}
  \caption{The correlation parameter as a function of radius at three times for a pair of fields in a four-field simulation.}
  \label{fig:covariance_radial}
\end{figure}

\cref{fig:cov_all} shows a weak dependence on the radius from which points are sampled. To investigate this dependence, we plot $\zeta$ in \cref{fig:covariance_radial} as a function of radius at the initial time and at 1.62 and 4.87 billion years. 
We observe that initially $\zeta$ is close to zero between $10\,\mathrm{kpc} < r < 15\,\mathrm{kpc}$ but it increases as the simulation runs. This is due to the fact that halos are initialized with eigenfunctions up to $20\,\mathrm{kpc}$ and have a smooth radially symmetric profile beyond this radius. This means that $\zeta = 1$ outside this region (see also Appendix \ref{app:ic}). As time progresses, some of this ``smoothness'' leaks from the region outside $20\,\mathrm{kpc}$ into the inner parts of the halo.

\begin{figure*}[tb]
    \includegraphics[width=\textwidth]{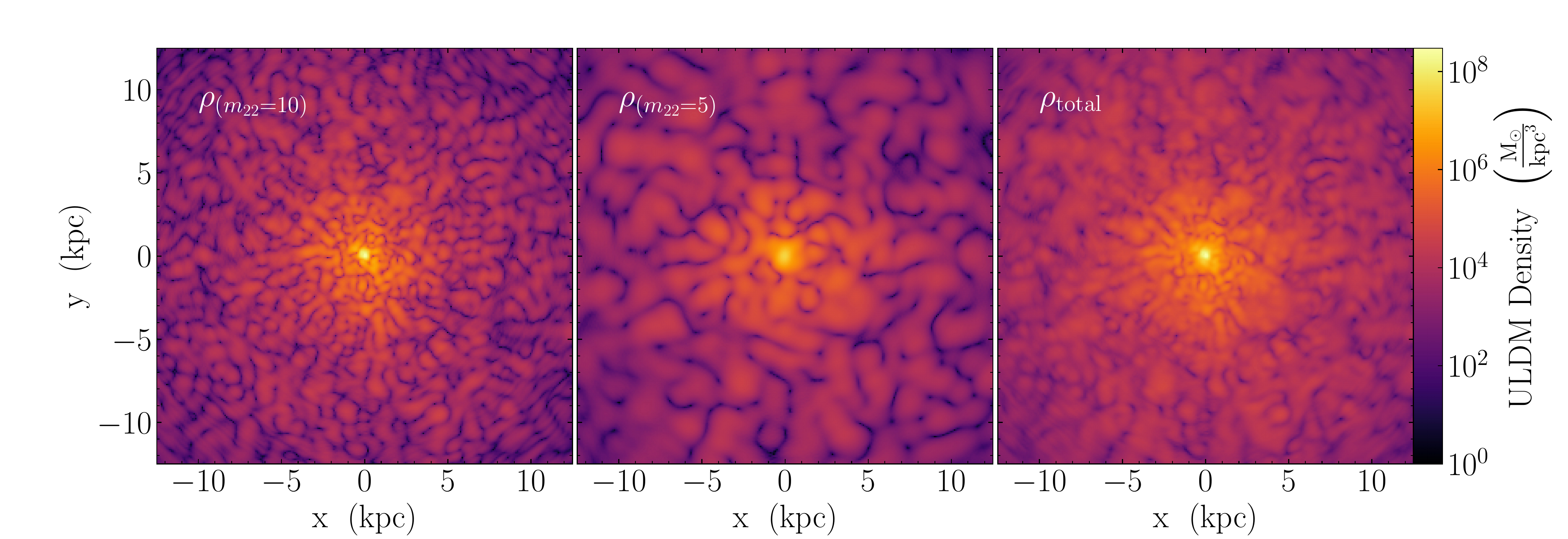}
    \caption{Densities of the two constituent fields in the two-field multi-mass scenario ($m_{22} =10$ left and $m_{22} = 5$ center), as well as the total density (right).}
    \label{fig:eigenIC}
\end{figure*}

\subsection{Multi-mass case}

\begin{figure}[tb]
  \includegraphics[width=0.5\textwidth]{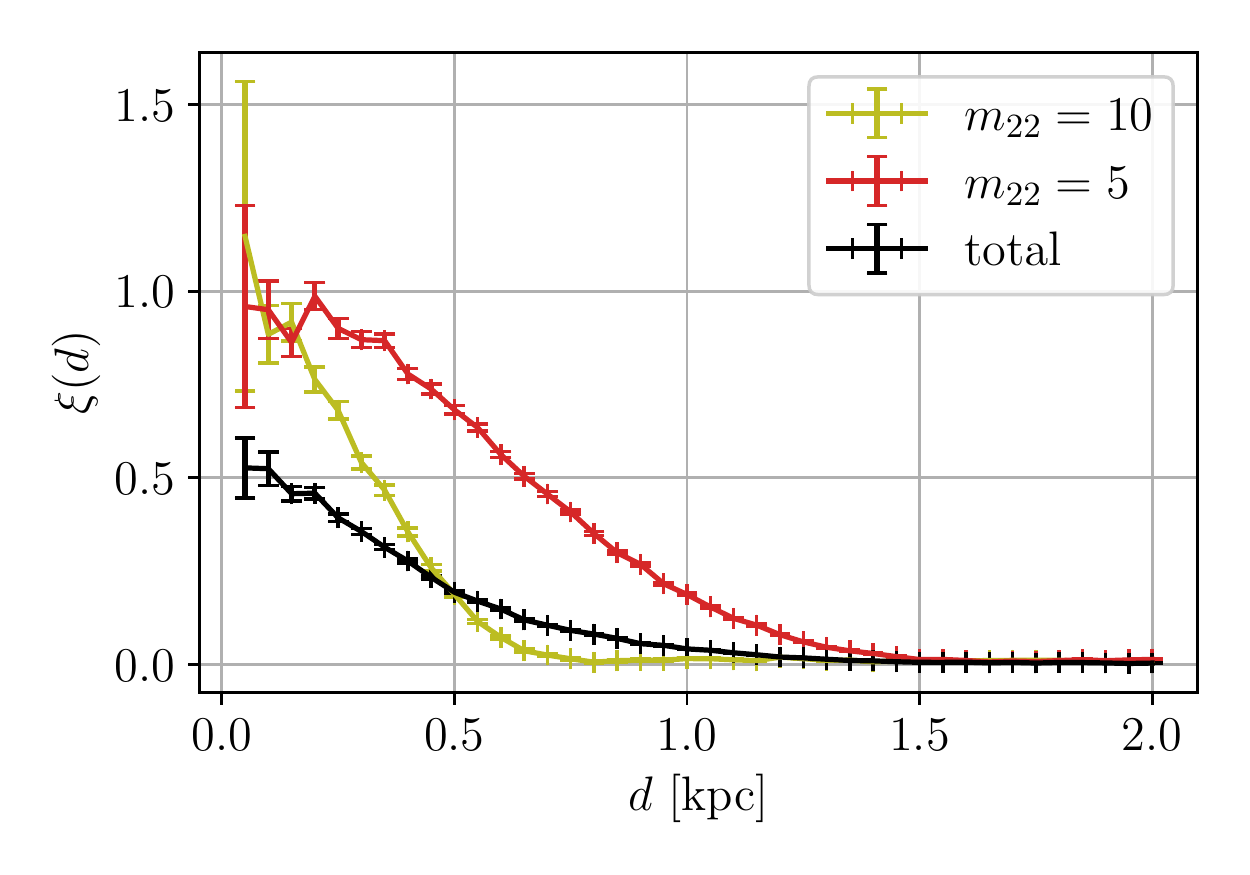}
  \caption{Two-point correlation function for the simulation of two fields with different masses. The combined density is smoother than each density separately.}
  \label{fig:two_mass_2ptcf}
\end{figure}

To investigate the smoothing of the granular structure with unequal masses we examine a representative two-field scenario, with  $m_{22} =10$ and $m_{22} =5$.  \cref{fig:eigenIC} shows the density  for both constituent fields and the total density at a representative time. The different de~Broglie wavelengths of the  fields are clearly visible and the combined density is again qualitatively smoother than either of the individual fields. This is made quantitative by the two-point correlation function which is again suppressed relative to that of the individual fields, as shown in \cref{fig:two_mass_2ptcf}. In addition, the granules in the heavy field are smaller so the two-point correlation function for this field is steeper and reaches zero at a smaller distance than that of the light field. 

This simulation verifies that the suppression of small-scale structure seen in the equal-mass case carries over to the multi-mass scenario. Clearly, this  may break down in the limit of an extreme mass ratio  -- if one field has a de~Broglie wavelength much larger than the others, it constitutes a smooth background relative to the structure present in the more massive fields. However, for mass differences of ${\cal O}(1)$, the qualitative dynamics do not appear to depend on whether we have strictly equal or merely similar masses. 

\section{Stellar dispersion constraints} \label{sec:stellarDispConstr}

The $\mathcal{O}(1)$ density fluctuations  resulting from the interference between different velocity streams in phase space 
can have an impact on the stellar dispersion of ultra-faint dwarf galaxies~\cite{Marsh:2018zyw, Dalal2022}. This is a result of stars experiencing an effective  headwind of granules as they orbit in the galactic potential which provides random kicks that increase
stellar velocity dispersion. Recently, this effect has been used to derive strong constraints on the ULDM mass for single-field scenarios. However, this effect is quadratically dependent on the amplitude of overdensities, i.e. $\Delta\sigma^2 \sim \delta\rho^2$ (see \cref{app:stellarDispersion}) so it is suppressed in a multifield ULDM scenario. 

\subsection{Equal-mass case}

We start by looking at scenarios with $N$ ultralight fields which all have approximately the same mass. We assume that the overdensities  are fully uncorrelated, ignoring the small but non-zero alignments found in our numerical simulations in the previous section.  As a star orbits in a halo it encounters granules from each of the constituent fields. Accelerations due to each granule add linearly, and we  treat the acceleration due to the granules in each field as independent encounters. The de~Broglie wavelength of each field is given by the collective macroscopic potential which is mostly unchanged relative to the single-field case. Therefore, we now have $N$ uncorrelated copies of the same granular  density field. This increases the number of encounters in the lifetime of the star, $n$, by a factor of $N$, i.e. $n \rightarrow n \, N$.

The total mass fraction in each field, indexed by $i$, is now $c_i = M_i/M$. The granule mass, therefore, decreases by the same factor, i.e. $\delta M \rightarrow \delta M(M_i/M)$, and as a result, the velocity kick from a granule encounter goes as $\delta v \rightarrow \delta v (M_i/M)$. Assuming that all fields contribute to the total mass equally, we can further deduce $\delta M \rightarrow \delta M / N$ and $\delta v \rightarrow \delta v /N$.   

The variances add linearly and so we can now write our constraint for the stellar dispersion as (see \cref{app:stellarDispersion} for details)
\begin{equation}
    \Delta \sigma_{\rm obs}^2 \ge \frac{n \, \delta v^2}{N} 
    \propto \frac{1}{Nm^3}\,,
\end{equation}
where $\Delta \sigma_{\rm obs}$ is the observed stellar dispersion. Consequently, for $N$ ultralight fields with the same mass and the same mass fraction, the impact of ULDM on stellar dispersions is relaxed by a factor of $N$ at a given mass. 

If the mass is not evenly distributed we obtain
\begin{equation}
\Delta \sigma_{\rm obs}^2 \ge n \, \delta v^2 \, \sum_i^N \left(\frac{M_i}{M}\right)^2  
    \propto \frac{1}{m^3}  \sum_i^N \left(\frac{M_i}{M}\right)^2\,.
\end{equation}

\subsection{Multi-mass case}

We calculate the more general multi-mass bound by noting that the variances again add linearly. The number of encounters $n_i$ with $i$-field granules and the amplitude of the gravitational kicks depends on the field masses, so we have 
\begin{equation}
\begin{split}
    \Delta \sigma_{\rm obs}^2 \ge \sum_i^N n_i \, \delta v_i^2 \left(\frac{M_i}{M}\right)^2 
    \propto \sum_i^N  \frac{1}{m_i^3}  \left(\frac{M_i}{M}\right)^2 \,. \label{eqn:multimassStellar}
\end{split}
\end{equation}
If the overall mass is evenly divided between the fields this simplifies to 
\begin{equation}
\Delta \sigma_{\rm obs}^2 \ge \frac{\sum_i^N n_i \, \delta v_i^2}{N^2} 
    \propto \frac{1}{N^2} \sum_i^N  \frac{1}{m_i^3}  \,. 
\end{equation}
The resulting stellar dispersion is a strong function of the mass spectrum. If the masses vary significantly, the term in the above sum corresponding to the lightest mass field, $m_L$, dominates the contribution to the velocity dispersion.  
Therefore, in the limit 
$\left(\frac{M_L}{M}\right)^2/m_L^3 \gg \left(\frac{M_i}{M}\right)^2/m_i^3$,
for all $i \ne L$, we find  
\begin{equation}
    \Delta \sigma_{\rm obs}^2 \ge \frac{1}{m_L^3} \left(\frac{M_L}{M}\right)^2 \label{eq:deltaunequal} \, .
\end{equation}

If, for example, each field has a roughly equal share of the total mass and the lightest mass is at least a few times lighter than the mass of any other field, i.e. $1/m_L^3 \gg 1/m_i^3$, stellar dispersion now scales as 
$1/(N^2 m_L^3)$. 
In contrast to the equal-mass, equal-mass-ratio case we now have an additional factor of $1/N$ because the lightest field has a dominant effect on the heating but only comprises $1/N$ of the total mass.

\section{Conclusions and Discussion} \label{sec:Conclusions}

We present simulations of multifield ULDM halos with a particular focus on how stellar heating constraints from ultrafaint dwarf galaxies are altered in the presence of more than one ULDM field.
We show that multifield models have smoother halos which 
reduce the extent to which orbiting stars are kicked by encounters with granules relative to the single-field case. We numerically evolve
multifield halos and compare our results to 
analytic approximation for
the heating dynamics. 
We perform the simulations with a modified version of \AxioNyxns, for up to four scalar gravitationally coupled fields. We examine one, two, or four equal-mass fields and two fields with masses varying by a factor of two.

As the number of constituent fields increases, the total density field in the halo becomes
smoother and the granular structure 
is washed out, even though each constituent field has 
the same 
amount
of granular structures as the single-field case.
We observe this for both the equal-mass and multi-mass cases. 
In the equal-mass case, the 2-point correlation function in the halo is proportional to $1/N$ and a consistent pattern is seen in the multi-mass case. This corresponds to the amplitude of the overdensity decreasing as $1/\sqrt{N}$.

Importantly, we verify that the suppression of fluctuations persists over time. The granules are effectively weakly-coupled oscillators, so it is conceivable that their motion could become synchronized, particularly in the equal-mass case. 
However, the correlation parameter $\zeta$ between the constituent fields remains small over cosmological timescales. 

Using analytic approximations we find that the expected stellar dispersion from gravitational heating of stars in ultra-faint dwarf galaxies scales as $1/(N m^3)$ for the equal-mass case, where $m$ is the mass of the ultralight particle.
Consequently, roughly 10 fields would be needed to relax a bound on the mass by a factor of two. In the multi-mass case, the heating strongly depends on how particle masses are distributed. In particular, it is dominated by the lowest-mass field if its particle mass $m_L$ is at least a few times lighter than that of any other field and $N$ is not very large. If all fields contain the same amount of the total mass, stellar dispersion then scales as $1/(N^2 m_L^3)$.

Formal observational bounds would require a more detailed analysis. Moreover, we note that many of the tightest bounds that have been proposed based on stellar dispersion rely on measurements of a small number of potentially idiosyncratic objects and the available dataset is likely to improve substantially in the coming years.  

Our work is based on halos constructed with eigenmodes and as the recent analysis by Huang {\em et al.}~\cite{Huang:2022ffc} demonstrates, in realistic cosmological multifield ULDM scenarios, not all of the constituent fields may form their own solitons. Moreover, in the limit that $N$ becomes  large, it is conceivable that none of the fields may form solitons and that the granule dynamics may also differ from the few-field case. This will be a worthwhile topic for future inquiry.  

Beyond the stellar dynamics, the central region of post-merger halos is of critical importance to the analyses of supermassive black hole merger dynamics. A single soliton undergoes both density oscillations and a random walk~\cite{Schive:2019rrw,Li2021}, potentially reheating an inspiraling supermassive black hole binary pair. We will address how the presence of multiple fields affects these phenomena in future studies.

It has been argued that the condensation of solitons is driven by the two-body relaxation of granular overdensities~\cite{Hui:2016ltb}. Therefore, it is possible the multifield picture would affect this timescale in a way similar to its impact on stellar dispersion. However, it has yet to be verified that this is the relevant timescale for soliton formation in realistic halo formation settings~\cite{Hui_2021}. Future studies of multifield structure formation, building on the recent work of Huang {\em et al.}~\cite{Huang:2022ffc} would allow this possibility to be assessed. Effects of multiple fields on the filaments or voids \cite{gallagher_coles_2022} present another possible research avenue. Separately, there is also a strong analogy between ULDM dynamics and the gravitational fragmentation of the inflaton condensate in the very early universe~\cite{Musoke:2019ima,Niemeyer:2019gab,Eggemeier:2020zeg,Eggemeier:2021smj,Eggemeier:2022gyo} and the dynamics of $N$-field ULDM may be mirrored in the primordial universe if inflation is driven by multiple fields~\cite{Dimopoulos:2005ac,Easther:2005zr,Dias:2015rca}. 

In summary, extending the single-field ULDM to multiple fields significantly alters the resulting cosmological dynamics. In particular, it reduces the amount of granularity in a dark matter halo which will in turn relaxes key observational constraints that are sensitive to the amplitudes of granules around the soliton, e.g. the heating of stellar orbits in ultrafaint dwarf galaxies. This possibility is actually more consistent with stringy arguments for the existence of very light axions -- given that they suggest the existence of many such fields rather than an isolated singlet -- and it has rich cosmological possibilities that are as-yet unexplored. 

\begin{acknowledgments}
We would like to thank Piotr T. Chru\'sciel, Oliver Hahn, Peter Hayman, Jens Niemeyer, and Nikhil Padmanabhan for useful discussions. Simulations presented here were performed on Vienna Scientific Cluster (VSC) (project number 71770). AE is supported by the U.S. Department of Energy under contract number DE-AC02-76SF00515. While at Yale, JZ was supported by the
Future Investigations in NASA Earth and Space Science and Technologies (FINESST) grant (award number 80NSSC20K1538). Research at Perimeter Institute is supported in part by the Government of Canada through the Department of Innovation, Science and Economic Development Canada and by the Province of Ontario through the Ministry of Colleges and Universities. RE, \textbf{}EK and YW acknowledge support from the Marsden Fund of the Royal Society of New Zealand. BE acknowledges support from the Deutsche Forschungsgemeinschaft.
\end{acknowledgments}

\appendix

\section{Eigenmodes for initial conditions}
\label{app:ic}
Here, we summarize the eigenfunction method for the construction of stable halos from Refs.~\cite{Yavetz:2021pbc, Zagorac:2021qxq}. We decompose the wavefunction of each field $\Psi$ into its orthogonal eigenmodes $\psi_j(\rb)$ which satisfy the time-independent Schr\"odinger equation,
\begin{equation}
\left( - \frac{\hbar}{2m} \nabla^2 + m \Phi \right) \psi_j = E_j \psi_j\,,
\end{equation}
where $E_j$ is the eigenenergy of that state. 
The total wavefunction is composed as 
\begin{equation}
\label{eq:waveeigenmodes}
    \Psi(\rb, t) = \sum_j a_j\psi_j(\rb) e^{-\ii E_j t/\hbar}\,,
\end{equation}
where $a_j$ is the amplitude of each eigenmode.
In our case, the time-dependent exponential factor can be omitted because we only construct the initial conditions and can therefore use $t=0$. This sum is truncated at the eigenmode whose energy corresponds to the energy of a particle on a circular orbit at the virial radius $r_{\rm vir}$.

We then factorize eigenfunctions into their radial and angular components,
\begin{equation}
\label{eq:RY}
    \psi_{j} (\rb) = \psi_{n\ell m}(r,\theta,\phi) = R_{n\ell}(r) Y_\ell^m (\theta, \phi)\,,
\end{equation}
where $ Y_\ell^m (\theta, \phi)$ are the spherical harmonics and the radial part $R_{n \ell}$ is obtained by solving 
\begin{equation}
\label{eq:radialRnl}
    -\frac{\hbar^2}{2m}\frac{d^2 u}{dr^2} +\left( \frac{\hbar^2}{2 m }\frac{\ell(\ell+1)}{r^2} +m \Phi(r) \right) u = E u
\end{equation}
with the new variable $u_{n \ell}(r) = rR_{n \ell}(r)$. 

We use the profile defined in \cref{eq:combined_profile} to solve the Poisson equation and determine $\Phi(r)$ which can then be used to solve \cref{eq:radialRnl} and to obtain the radial part of the eigenmodes. This is done by discretizing the radial domain and rewriting \cref{eq:radialRnl} in a matrix form. We find the eigenfunctions of this matrix with a tridiagonal matrix solver from {\sc scipy.linalg}.

Once we have the eigenmodes, we can create a constructed density using $\rho_{\rm con}(\rb) = m |\Psi(\rb)|^2$.
Combining \cref{eq:waveeigenmodes,eq:RY}, as well as using $\sum_m |Y_\ell^m(\theta, \phi)|^2 = (2\ell+1)/4\pi$, the radial profile of the constructed density becomes
\begin{equation}
    \rho_{\rm con}(r) =\frac{1}{4\pi} \sum_{n\ell} (2\ell+1) |a_{n\ell}|^2 |R_{n\ell}(r)|^2\,.
\end{equation}
The coefficients $a_{n\ell}$ which determine the amplitudes of each eigenmode are independent of the magnetic number $m$. To determine their value we need to minimize the cost function
\begin{equation}
    C (\rho_{\rm tar}, \rho_{\rm con})= \frac{1}{r_{\rm fit}} \int_0^{r_{\rm fit}} \mathrm{d}r \left( \frac{\rho_{\rm con} - \rho_{\rm tar}}{\rho_{\rm tar}}\right)^2\,.
\end{equation}
This is done numerically using the optimization methods from \textsc{scipy}.
Once the amplitudes of each eigenmode are determined, the three-dimensional realization of a ULDM halo is obtained by performing the sum 
\begin{equation}
        \Psi(r, \theta, \phi) = \sum_{n\ell} a_{n\ell} R_{n\ell}(r) Y_\ell^m (\theta, \phi) \: e^{\ii f_{n\ell m}}
\end{equation}
in each point $(r, \theta, \phi)$. Furthermore, each eigenmode is multiplied by a random phase $ f_{n\ell m}$ whose value is between $0$ and $2\pi$. This ensures that the constructed wave function has an imaginary as well as a real part and that the halo is stable. 

To speed up this construction, we use the so-called isotropic fit described in Ref.~\cite{Yavetz:2021pbc} which means that the eigenfunctions are binned by similar energy eigenvalues and all eigenfunctions in the same bin have the same $a_{n\ell}$.

We initialize the halo up to the virial radius $r_{\rm vir} = 20\, {\rm kpc}$. Outside of this radius, the halo has a smooth suppressed radially-symmetric NFW profile which ensures that periodic boundary conditions do not affect the dynamics of the halo. There is, however, quite a sharp transition between the eigenfunctions region and the smooth region which can produce some spurious effects. This method could be further improved by smoothing this sharp transition in some way, but for our purpose, we avoid spurious effects by analyzing a spherical region within only $\sim 3/4$ of $r_{\rm vir}$.    

\section{Stellar dispersion constraints in the one field case} \label{app:stellarDispersion}

Here we discuss the central argument of stellar heating for a single classical field. Interference between streams results in the density field being broken into granules with $\mathcal{O}(1)$ oscillations in the density. 
On the de Broglie scale, the fluctuations in the density are proportional to the average density at any given point $\delta \rho \sim \rho$, 
when sufficiently far from the central soliton that gravitational heating is dominated by particles encountering granules rather than the dynamics of the soliton. 

The density field of the dark matter is composed
of a collection of such granules. As a star travels in the vicinity of one of these granules it receives an acceleration due to the potential gradient of the overdensity. In the weak deflection limit, this alters the velocity as~\cite{Binney_2008}
\begin{align}
    \delta v = \frac{2 \, G \, \delta M}{r \, \sigma_{\rm DM}} \,,
\end{align}
where $r$ is the impact parameter and assumed to be approximately the de~Broglie wavelength associated with the granule of size $r\sim \lambda = \hbar / \sigma_{\rm DM} / m$, $\delta M$ is the total mass of the dark matter granule $\delta M \sim \delta\rho \, r^3$, and $\sigma_{\rm DM}$ is the dark matter velocity dispersion. 

The number of encounters in the lifetime of the star, $n$, is given by comparing the granule crossing time $r/\sigma_{\rm DM}$ to the total integration time $t$, resulting in $n \sim \sigma_{\rm DM} t / r$. As each encounter kicks the star randomly, we can add the variance of the kicks linearly and interpret the sum as the predicted impact of granules on the velocity dispersion of stars due to the dark matter for a specific model, i.e. 
\begin{align}
    \Delta \sigma_\mathrm{pred}^2 \sim n \, \delta v^2 \, .
\end{align}
If this is larger than the observed stellar dispersion relation $\Delta \sigma_{\rm obs}^2$ then the considered dark matter model can be ruled out. 
Using that the stellar dispersion $\sigma_*$ can be related to the enclosed mass at the half-light radius $r_{1/2}$ by $M_{1/2} \approx 3 \sigma_*^2 r_{1/2}/G$ and that the density is approximately $\rho = M_{1/2} / r_{1/2}^3$, we can rewrite the bounds on the dark matter mass as
\begin{align} \label{eqn:bounds}
    \Delta \sigma_{\rm obs}^2 \ge n \, \delta v^2 =  9 \left(\frac{\sigma_*}{\sigma_{\rm DM}}\right)^4 \left( \frac{\hbar}{m} \right)^3 \frac{t}{r^4_{1/2}} \,,
\end{align}
recovering the result in Ref.~\cite{Dalal2022}.

\bibliography{bibliography}

\end{document}